# Title
*Global daily 1km land surface precipitation based on cloud cover-informed downscaling*

## Authors
Dirk Nikolaus Karger[1,2,3], Adam M. Wilson[4], Colin Mahony[3], Niklaus E. Zimmermann[1], Walter Jetz[2,3]

## Affiliations
1. Swiss Federal Research Institute for Forest, Snow, and Landscape Research (WSL), Zürcherstrasse 111, 8903 Birmensdorf, Switzerland
2. Department of Ecology and Evolutionary Biology, Yale University, 165 Prospect Street, New Haven, CT 06520-8106, USA
3. Center for Biodiversity and Global Change, Yale University, 165 Prospect Street, New Haven, CT 06520-8106, USA
4. Department of Geography, University at Buffalo, 120 Wilkeson Quad, Buffalo NY, 14261, USA

corresponding author(s): Dirk Nikolaus Karger (dirk.karger@wsl.ch), Walter Jetz (walter.jetz@yale.edu)

## Abstract
High-resolution climatic data are essential to many questions and applications in environmental research and ecology. Here we develop and implement a new semi-mechanistic downscaling approach for daily precipitation estimate that incorporates high resolution (30 arcsec, ≈1km) satellite-derived cloud frequency. The downscaling algorithm incorporates orographic predictors such as wind fields, valley exposition, and boundary layer height, with a subsequent bias correction. We apply the method to the ERA5 precipitation archive and MODIS monthly cloud cover frequency to develop a daily gridded precipitation time series in 1km resolution for the years 2003 onward. Comparison of the predictions with existing gridded products and station data from the Global Historical Climate Network indicates an improvement in the spatio-temporal performance of the downscaled data in predicting precipitation. Regional scrutiny of the cloud cover correction from the continental United States further indicates that CHELSA-EarthEnv performs well in comparison to other precipitation products. The CHELSA-EarthEnv daily precipitation product improves the temporal accuracy compared with a large improvement in the spatial accuracy especially in complex terrain.

## Background & Summary
High resolution information on precipitation is essential in many scientific fields, ranging from ecology, agriculture, forestry, to global change impact studies [1–3]. Spatiotemporal precipitation data is usually derived from a range of different sources, including satellites, reanalysis, global circulation models, or precipitation gauges [4,5]. However, each of these sources on their own have limitations in coverage, accuracy, or detail, impeding many downstream uses, especially those addressing large spatial and temporal extents [6,7].

Reanalysis data products such as ERA-5 [8,9], MERRA 2 [10,11] or MSWEP [12] overcome these constraints by combining data from a variety of sources. To date, however, they remain limited to rather coarse spatial resolutions such as 0.5° ~0.25°, i.e. ca. 55-27km near the equator. This is much coarser than the scale of many environmental and ecological processes and the associated data requirements for ecosystem management and conservation. This resolution is furthermore too coarse to capture orographic precipitation in complex terrain [13–15]. Global circulation and weather models such as WRF-ARF [16] and ICON [17,18] are able to run at high spatial resolutions of 1km, but are still heavily constrained by computational limits [7]. Currently, global kilometer scale models are only able to archive a simulation throughput of 0.043 SYPD (Simulated years per day) [19], which amounts to an 100x shortfall compared to computationally efficient simulations defined as 1 SYPD [7,20]. Even with the largest supercomputers and with state of the art climate models, as well as large financial investments, this shortfall can only be reduced to approximately 20x [21].

Although achieving 1km resolution in numerical climate models is important to quantify effects such as deep convection or surface drag [21], studies focusing on the impact of climate on different systems often rely on a limited set of climatic variables. In ecological studies for example, precipitation together with minimum-, mean-, and maximum temperatures, are often used to delineate occurrences of species [22]. It is common to characterize the range of a species by its climatic envelope in e.g. species distribution models (SDMs) using a rather simple set of climatic predictors and derivations thereof [23,24]. This means that for applications like these, only a subset of available variables needs to be downscaled to finer spatial resolution. Environmental scientists or climate impact modelers in need of high-resolution precipitation data, therefore often resort to data from computationally less intensive methods. One such method is the spatial interpolation of data from climatic stations. Here, precipitation gauges form the input for interpolation [25] or regression models to achieve a high spatial resolution either with or without additional, often terrain derived, predictors [26–28]. Such interpolations however, usually suffer from a spatially uneven station density [29–32] and severely underestimate snowfall [6,33–35]. While gauge undercatch can be corrected using statistical methods in combination with steam flow observations [36], the spatially uneven distribution of gauges can lead to false parametrizations of precipitation lapse rates in regression based interpolation methods [37]. One method to overcome the limitation imposed by uneven station density is to directly downscale the output of reanalysis data calculated at coarser cell size [37–39]. However, there are still interpolations and parametrizations involved which account for processes not resolved at the original model resolution.

This uneven distribution of gauges can be overcome by the use of satellite data [2,40–43], which offers spatially more complete information of precipitation patterns. Yet, satellites also detect snowfall poorly [44,45], meaning that the satellite-derived amounts of precipitation have to be corrected. This is usually done by bias-correction using station observations [46–50]. Although satellite precipitation products generally have a higher horizontal resolution than reanalysis data, they are globally still not available at resolutions of 1km needed for local impact studies. However, available at this very high resolution is cloud cover from satellites, which can potentially lead to improved spatial representation of precipitation. The established relationship between cloud occurrences and precipitation [51,52] scales precipitation with cloud cover frequency such that if no clouds occur there is no precipitation, and increasing cloud frequency translates to increasing precipitation [53].

Here we merge data from a downscaled reanalysis (ERA5) using the CHELSA algorithm [37,39] with cloud cover information derived from MODIS from the EarthEnv layer suite (https://www.earthenv.org) [54] to achieve a better representation of the fine-scale variation of global precipitation patterns. The presented CHELSA-EarthEnv daily precipitation data at ~1km horizontal resolution offers a more reliable characterization of precipitation in topographically heterogeneous regions and supports a range of applications that require high resolution precipitation data.

## Methods

**Bias correction of ERA5 precipitation data**

ERA5 shows an increased performance over its predecessor ERA-Interim in several attributes[55] and especially in precipitation [56]. Nevertheless, for application in impact studies, there is often still a significant bias observed in several parameters that need to be accounted for[57]. For accumulated parameters such as total daily precipitation, we used the monthly sum of the hourly precipitation from ERA5 $p_{era}$ to assess this bias. ERA5 generated estimates of the surface precipitation, similar to its predecessor ERA-Interim are extracted from short range forecasts, which vary considerably with forecast length [58]. This bias in the short-range forecasts can be a problem for monthly and climatological means as it accumulates over time [58]. Several methods exist to account for such biases but most of them require gapless gridded observational data which comes with an inherent interpolation error itself [57]. To correct the bias in the ERA5 precipitation estimates and account for the interpolation errors we therefore performed a bias correction which consists of three steps.

1. One very common approach to account for reanalysis bias is to calculate the difference between baseline precipitation from the reanalysis and the observed precipitation from station data and apply this 'change factor' to the reanalysis data. We apply a monthly bias correction on the accumulated ERA5 precipitation for each month $p_{sim}$. We used the monthly accumulated precipitation $p_{obs}$ of the gridded GPCC 2018 dataset [59]. The bias correction in earlier versions of CHELSA did not adequately interpolate across the dateline, which caused artefacts in this region. To correct for this we reprojected both $p_{era}$ and $p_{obs}$ to a North Pole Azimuthal Equidistant projection (EPSG: 102016) to allow interpolation across the dateline. We then calculate the monthly bias $R_m$ caused by the ERA5 parametrization, and the excessive or insufficient precipitation of the forecast algorithm for each month using:

$$R_m^{obs} = \frac{p_{obs} + c}{p_{sim} + c}$$

with $c$ being a constant of 0.0001 kg*m$^{-2}$*s$^{-1}$ to avoid division by zero. We only used grid cells with meteorological stations present for the calculation of the observed bias $R_m^{obs}$. The forecast algorithm used to produce the precipitation amounts for ERA5 exhibits a considerable bias (too much or too less precipitation), that has a coherent spatial structure, with a larger bias over high elevation terrain, or specific landforms such as tropical rainforests. Based on this observation, we assumed that grid cells without stations share a similar bias as their neighbouring stations.

2. To achieve a gap-free bias correction grid surface, we interpolated the gaps in the $R_m$ grid using a multilevel B-spline interpolation[60] with 14 error levels optimized using B-spline refinement to a 0.25° resolution. The multilevel B-spline approximation [60] applies a B-spline approximation to $R_m$ *starting* with the coarsest lattice $\phi_0$ from a set of control lattices $\phi_0, \phi_1, ..., \phi_n$ with $n$ = 14 that have been generated using optimized B-spline refinement [61]. The resulting B-spline function $f_0(R_m^{obs})$ gives the first approximation of $R_m$. $f_0(R_m)$ leaves a deviation between $\Delta^1 R_m^{obs}{}_c = R_m^{obs} - f_0(x_c, y_c)$ at each location $(x_c, y_c, R_m^{obs}{}_c)$ [61]. Then the next control lattice $\phi_1$ is used to approximate $f_1(\Delta^1 R_m^{obs}{}_c)$ [61]. Approximation is then repeated on the sum of $f_0 + f_1 = R_m^{obs} - f_0(x_c, y_c) - f_1(x_c, y_c)$ at each point $(x_c, y_c, R_m^{obs}{}_c)$ $n$ times resulting in the gap free interpolated bias surface $R_m^{int}$ [61].

3. The bias correction surface $R_m^{int}$ is then multiplied with the ERA5 precipitation $p_{sim}$ to get the bias corrected monthly precipitation sums $p_m^{cor}$ at 0.25° resolution:

$$p_m^{cor} = p_{sim} * R_m^{int}$$

**Orographic wind effects**

Orographic effects are among the most reported drivers of precipitation [62–66]. Orographic effects have been taken into account using a variant of the CHELSA V1.2 algorithm which uses a parametrization of orographic rainfall based on wind fields [67–70]. We used daily *u*-wind and *v*-wind components at the 10-m level of ERA5 as underlying wind components. As the calculation of a windward leeward index $H$ (hereafter: wind effect) requires a projected coordinate system, both wind components (u-wind, v-wind) were projected to a world Mercator projection and then interpolated to a 3 km grid resolution using a multilevel B-spline interpolation similar to the one used for the bias correction surface. The resolution of 3km was chosen as resolutions of around 1km would over-represent orographic terrain effects [26]. The wind effect *H* was then calculated by multiplying the windward $H_w$ and leeward $H_L$ components calculated using:

$$H_W = \frac{\sum_{i=1}^{n} \frac{1}{d_{WHi}} tan^{-1}\left(\frac{d_{WZi}}{d_{WHi}^{0.5}}\right)}{\sum_{i=1}^{n} \frac{1}{d_{LHi}}} + \frac{\sum_{i=1}^{n} \frac{1}{d_{LHi}} tan^{-1}\left(\frac{d_{LZi}}{d_{LHi}^{0.5}}\right)}{\sum_{i=1}^{n} \frac{1}{d_{LHi}}}$$

$$H_L \frac{\sum_{i=1}^{n} \frac{1}{\ln(d_{WHi})} tan^{-1}\left(\frac{d_{LZi}}{d_{WHi}^{0.5}}\right)}{\sum_{i=1}^{n} \frac{1}{\ln(d_{LHi})}}$$

where $d_{WHi}$ and $d_{LHi}$ refer to the horizontal distances between the focal 3km grid cell in windward and leeward direction and $d_{WZi}$ and $d_{LZi}$ are the corresponding vertical distances compared with the focal 3km cell following the wind trajectory. Distances are summed over a search distance of 75 kilometers as orographic airflows are limited to horizontal extents between 50 – 100 km [71,72]. The second summand in the equation for $H_{W,L}$ where $d_{LHi} < 0$ accounts for the leeward impact of previously traversed mountain chains. The horizontal distances in the equation for $H_{W,L}$ where $d_{LHi} \geq 0$ lead to a longer-distance impact of leeward rain shadow. The final wind-effect parameter, which is assumed to be related to the interaction of the large-scale wind field and the local-scale precipitation characteristics, is calculated as:

$$H = H_{W,L} \rightarrow d_{LHi} < 0 * H_{W,L} \rightarrow d_{LHi} \geq 0$$

and generally, takes values between 0.7 for leeward and 1.3 for windward positions. Both equations were applied to each grid cell at the 3km resolution in a World Mercator projection.

We used the boundary layer height $PBL$ from ERA5 as an indicator of the pressure level that has the highest contribution to the wind effect. *PBL* and *H* have been interpolated to a 30 arc second using a B-spline interpolation. To create a boundary layer height corrected wind effect $H_B$, the wind effect grid $H$ containing was then proportionally distributed to all grid cells falling within a respective 0.25° grid cell using:

$$H_B = \frac{H}{1 - \left(\frac{|z - PBL_z| - z_{max}}{h}\right)}$$

with $z_{max}$ being the maximum distance between the boundary layer height $B_z$ *at elevation z* and all grid cells at a 30 arc sec resolution falling within a respective 0.25° grid cell, $h$ being a constant of 9000 m, and $z$ being the respective elevation from the Global Multi-resolution Terrain Elevation Data (GMTED2010) [72] with:

$$PBL_z = PBL + z_{ERA} + f$$

where $B$ is the height of the daily means of the boundary layer from ERA5, and $z_{ERA}$ is the elevation of the ERA5 grid cell. The boundary layer height provided by ECMWF is based on the Richardson number [73] which is usually at the lower end of the elevational spectrum compared to other methods [74]. We therefore tuned our model by adding a constant of 500 m similar to the approach in the original CHELSA algorithm [37].

Although the wind effect algorithm can distinguish between the windward and leeward sites of an orographic barrier, it cannot distinguish extremely isolated valleys in high mountain areas [75]. Such dry valleys are situated in areas where the wet air masses flow over an orographic barrier and are prevented from flowing into deep valleys [75]. These effects are however mainly confined to large mountain ranges, and are not as prominent in intermediate mountain ranges [72]. To account for these effects, we used a variant of the windward-leeward equations with a linear search distance of 300 km in steps of 5° from 0° to 355° circular for each grid cell. The calculated leeward index was then scaled towards higher elevations using:

$$E = \left(\frac{\sum_{i=1}^{n} \frac{1}{\ln(d_{WHi})} tan^{-1}\left(\frac{d_{LZi}}{d_{WHi}^{0.5}}\right)}{\sum_{i=1}^{n} \frac{1}{\ln(d_{LHi})}}\right)^{\frac{z}{h}}$$

which rescales the strength of the exposition index relative to elevation $z$ from GMTED2010, and gives valleys at high elevations larger wind isolations $E$ than valleys located at low elevations. The correction constant $h$ was set to 9000 m to include all possible elevations of the DEM and because values of $z > h$ could otherwise lead to a reverse relationship between $z$ and $E$.

$$p_{I_c} = E * H_B$$

will give the first approximation of precipitation intensity $p_{I_c}$ at each grid location $(x_c, y_c)$.

**Precipitation including orographic effects**

To achieve the distribution of daily precipitation $p_o$ given the approximated precipitation intensity $p_{I_c}$ at each grid location $(x_c, y_c)$, we used a linear relationship between $p_m^{cor}$ and $p_{I_c}$ using:

$$p_o = \frac{p_{I_c}}{\frac{1}{n}\sum_{i=1}^{n} p_{I_{c_i}}} * p_m^{cor}$$

where $n$ equals the number of 30 arc sec. grid cells that fall within a 0.25 grid cell. This equation ensures that the precipitation at 0.25° resolution exactly matches the mean precipitation of all 0.0083334° cells that overlap with a 0.25° cell.

The GPCC dataset used for the bias correction does not include a correction for gauge undercatch. We therefore additionally correct for gauge undercatch using a downscaled version of the bias correction layers from Beck et. al 2020 [36]. We downscaled the bias correction surfaces to 0.0083334° by using a moving window regression with a search radius of three cells and elevation from GMTED2010 as predictor. We then multiplied this downscaled bias correction layer with the $p_o$.

**Monthly cloud frequencies**

To derive monthly cloud frequencies, we used the internal cloud mask in the PGE11 program that relies on two reflective and one thermal test MODIS MOD09 atmospherically corrected surface reflectance product [76,77]. The reflective tests include the shortwave and middle infrared data combined in the "middle infrared anomaly" index (MIRA = ρ20,21 − 0.82ρ7 + 0.32ρ6, where ρ indicates MODIS band number). The second test uses reflectance at 1.38 microns (1.38 mic = ρ26). The MIRA and the 1.38 mic reflectance are designed to be complementary, with MIRA efficiently detecting low or high reflective clouds [77], while 1.38 mic effectively detects high (and potentially not very reflective) clouds. Additionally, a thermal test is used to identify pixels with high infrared reflectance anomalies (e.g., fires, sun-glint, and high albedo surfaces) with respect to near-surface (2 m) air temperature computed by the NCEP reanalysis model [78]. The MOD09 cloud algorithm was designed to minimize confusion over snow and ice by taking the surface air temperature into account. Like many cloud masks, the MOD09 detection algorithm has a binary response (cloudy/not cloudy) and does not retain an estimate of confidence in cloud state (i.e., probability that the pixel was actually cloudy given the tests). We extracted the daily cloud flags from bit 10 of the daily daytime surface reflectance product "state 1 km" Scientific Data Set (SDS) from both the Terra (MOD09GA, collected at approximately 10:30 AM local time) and Aqua (MYD09GA, approximately 1:30 PM) satellites. The time series of monthly cloud frequencies (proportion of days with a positive cloud flag) was calculated separately for the daily MOD09GA and MYD09GA data using the Google Earth Engine application programming interface (http://earthengine.google.org/).

**Cloud frequency correction of daily precipitation estimates**

We include monthly cloud frequencies $cf_m$ into the daily precipitation estimates assuming that the frequency of cloud occurrences is related to precipitation events and their geographic distribution carries a spatial signal of precipitation [51,52]. Strictly we assume that where no clouds occur, no precipitation occurs, and where clouds occur more frequently, more precipitation occurs [53]. To achieve the distribution of daily precipitation $p$ given the approximated orographic corrected precipitation $p_o$ at each grid location $(x_c, y_c)$, we first approximate the cloud cover corrected precipitation intensity using:

$$p_{cf_c} = p_o * cf_m$$

This however distorts the precipitation amount of each grid cell. We therefore repeat the step used to create orographic precipitation in a similar manner by estimating daily precipitation $p$ at each grid location $(x_c, y_c)$ using:

$$p = \frac{p_{cf_c}}{\frac{1}{n}\sum_{i=1}^{n} p_{cf_{c_i}}} * p_m^{cor}$$

where $n$ equals the number of 30 arc sec. grid cells that fall within a 0.25 grid cell.

## Data Records

The dataset [79] is available at EarthEnv (https://doi.org/10079/MOL/6f52b80d-0a41-40f7-84ec-873458ca6ee6). All files are provided as georeferenced tiff files (GeoTIFF). GeoTIFF is a public domain metadata standard which allows georeferencing information to be embedded within a TIFF file. Additional information included in the file are: map projection, coordinate systems, ellipsoids, datums, and fill values.

GeoTIFF can be viewed using standard GIS software such as:
SAGA GIS – (free) http://www.saga-gis.org/
ArcGIS - https://www.arcgis.com/
QGIS - (free) www.qgis.org
DIVA – GIS - (free) http://www.diva-gis.org/
GRASS – GIS - (free) https://grass.osgeo.org/

All files contain variables that define the dimensions of longitude and latitude (Table 1). The time variable is usually encoded in the filename.
All files are in a geographic coordinate system referenced to the WGS 84 horizontal datum, with the horizontal coordinates expressed in decimal degrees. The extent (minimum and maximum latitude and longitude) are a result of the coordinate system inherited from the 1-arc-second GMTED2010 data which itself inherited the grid extent from the 1-arc-second SRTM data.

**Table 1** Grid extent and resolution of the GeoTIFF files:

| item | value |
|---|---|
| Resolution | 0.0083333333 |
| West extent (minimum X-coordinate, longitude): | −180.0001388888 |

| South extent (minimum Y-coordinate, latitude) | −90.0001388888 |
| East extent (maximum X-coordinate, longitude) | 179.9998611111 |
| North extent (maximum Y-coordinate, latitude) | 83.9998611111 |
| Rows | 20,800 |
| Columns | 43,200 |

The filename includes the respective model used, the variable short name, the respective time variables, and the version of the data:

[Model]_[short_name]_[day]_[month]_[year]_[Version].tif

There are two different models available. CHELSA which includes the results from the bias correction and orographic correction, and CHELSA_EarthEnv which includes the cloud cover correction as well.

The unit of the precipitation is CHELSA_EarthEnv is: (kg*m^-2*day^-1)/100

## Technical Validation

To validate the performance of CHELSA_EarthEnv we are focusing on (a) the downscaling performance by calculating different performance metrics between coarse and high resolution and comparing observations from meteorological stations and (b) a comparison with similar high-resolution precipitation datasets (Table 2) within the continental United States where meteorological station density is high and of good quality.

**Validating the downscaling performance**

To validate if the downscaling to 0.0083334° resolution leads to a better performance over the coarser 0.25° gridded data that was used as forcing, we compare both resolutions with precipitation measured at Global Historical Climate Network – daily weather stations (GHCN-D)[80]. The 0.25° resolution has been chosen as benchmark as it is the resolution of the forcing ERA5 data that is used as an input for CHELSA_EarthEnv. To set the performance changes in to four comparable products (Table 2): PRISM AN81d, MSWEP 2.1, CHIRPS 2.0, and WorldClim 2.1 and repeated the analysis with these datasets over the continental United States except Alaska.

**Table 2.** Overview of the precipitation datasets used for comparison and their respective properties and methodologies.

| model | version | Native spatial resolution | temporal resolution | precipitation source data | method | citation |
|---|---|---|---|---|---|---|

| PRISM | AN81d | 0.041667° | daily | point rain gauge station data | constrained regression & interpolation | 26 |
| MSWEP | 2.1 | 0.1° | 3 hourly | point rain gauge station data, reanalysis data, satellite observations | multi-source weighted ensemble | 5 |
| CHIRPS | 2.0 | 0.05° | daily | satellite observations, point rain gauge station data | modified inverse distance weighting | 43 |
| CHELSA | 2.1 | 0.0083334° | daily | reanalysis data, gridded rain gauge station data | model output statistics | - |
| WorldClim | 2.1 | 0.041667° | monthly | point rain gauge station data | regression & interpolation | 81 |

**Accessing the global downscaling performance across several metrics**

To validate the performance of CHELSA_EarthEnv globally we compare it to observations at metrological stations from the GHCN-D [80] network for the time 2003-2016. We use only stations without any quality flags and compare them to the precipitation data at the coarse 0.25°, and the high 0.0083334° spatial resolution.

Downscaling can affect different aspects of model performance such as bias, variability, or correlation coefficients. To test in a first step which metric is affected by the applied downscaling we calculated for each grid cell separately the Kling-Gupta efficiency (KGE) scores from daily time series from 2003 to 2016. KGE is a performance metric combining correlation, bias, and variability [82,83] and is defined as follows:

$$KGE = 1 - \sqrt{(r-1)^2 + (\beta-1)^2 + (\gamma-1)^2}$$

where the correlation component *r* is represented by the Pearson's correlation coefficient, the bias component *β* by the ratio of estimated and observed means, and the variability component *γ* by the ratio of the estimated and observed coefficients of variation:

$$\beta = \frac{\mu_s}{\mu_s} \; and \; \gamma = \frac{\frac{\sigma_s}{\mu_s}}{\frac{\sigma_o}{\mu_o}}$$

where *μ* is the mean and *σ* the standard deviation, and the subscripts *s* and *o* indicate simulated and observed, respectively. KGE, *r*, *β*, and *γ* values all have their optimum at 1. KGE values between -0.41 and 1 indicate that the model estimates precipitation better than just taking the mean of the recorded precipitation at the gauges [84].

We also calculated the percent bias (*pbias*) that reflects the average tendency of the modelled precipitation values $p_{sim}$ to be larger or smaller than their observed values $p_{obs}$ at the stations. The optimal value of *pbias* is 0, with low values indicating accurate model simulation. Positive values indicate an overestimation, whereas negative values indicate an underestimation. *pbias* is defined as follows:

$$pbias = 100 * \left( \frac{\sum_{i=0}^{n}(p_{sim_i} - p_{obs_i})}{\sum_{i=0}^{n} p_{obs_i}} \right)$$

Additionally, we also report the mean absolute error (*mae*) which is defined as:

$$mae = \frac{1}{n} \left( \sum_{i=0}^{n} |p_{sim_i} - p_{obs_i}| \right)$$

and the root mean squared (*rmse*) error which is defined as:

$$rmse = \sqrt{\frac{1}{n} \left( \sum_{i=0}^{n} (p_{sim_i} - p_{obs_i})^2 \right)}$$

**Accessing the regional performance**

To compare the results to similar precipitation datasets, we use GHCN-D and four other gridded datasets (Table 2) that provide data over the same time period: PRISM (AN81d) [26], MSWEP 2.1 [12], and CHIRPS 2.0 [43], and WorldClim 2.1 [81]. PRISM is a high-resolution precipitation dataset for the United States that, similar to CHELSA_EarthEnv, takes orographic effects into account and additionally profits from a dense quality-controlled network of weather stations. While PRISM uses a regression approach to predict long term precipitation climatologies, daily precipitation is derived from climatologically aided interpolation (CAI) [85]. MSWEP 2.1 is a merged product from various sources (weather stations, reanalysis data, satellite observations) and consistently has high performance scores in comparison to other precipitation products [6]. CHIRPS is a high-resolution precipitation dataset, that integrates remote sensed precipitation with observations from weather stations. Additionally, we also include the WorldClim 2.1 data in our comparison. Although WorldClim 2.1 does not offer daily data, it provides monthly timeseries that has been created using climatologically aided interpolation of the CRU-TS 4.03 data [86]. All these datasets have been aggregated over the period 2003-2016 to annual means to gain a comparable temporal extent as CHELSA_EarthEnv. We then compare these data from the different datasets with observations from GHCN-D [80] for the continental United states except Alaska. Within this spatial extent all five products overlap and the quality of the stations can be considered as high. All products have additionally been aggregated to a 0.25° grid resolution by taking the mean of all grid cells overlapping with a 0.25° grid cell in WGS84 geographic projection. We then used all stations with data available between 2003 and 2016 and without any quality flag (58,071 stations) and extracted precipitation from both the highest available spatial resolution of the different datasets (Table 2) and the coarse 0.25° resolution using a nearest neighbour approach. We then calculated the differences in absolute bias between coarse and high resolution, and compared these among products using an ANOVA with post-hoc Tukey HSD test.

**Comparison with PRISM**

The validation of the temporal accuracy done using the GHCN-D station data gives information how well a product reproduces precipitation directly at the locations of these stations. All products we compare to CHELSA here are however, at least partly, parameterized on a subset of the GHCN-D stations as well. This often leads to a high fit with station data in all products that use exactly these climate stations at the locations of the stations. However, predicted precipitation patterns between stations, where the data is actually interpolated or predicted cannot be validated in this way. The performance of a model to predict the spatial patterns of precipitation correctly could for example be accessed by a cross validation approach, but this is not possible without the station data or the source code of the respective model being available. As the exact station data each dataset uses are generally not available, we use the spatially explicit PRISM model as a benchmark for comparison. PRISM has a very high accuracy and captures small scale precipitation gradients well. It uses the highest amount of meteorological stations of all models compared here. It is however, also a model and therefore has its own inherent biases. To compare models, we aggregated the daily values (monthly for WorldClim) over 2003-2016 to mean annual precipitation, and calculated the bias and correlation between products.

**Comparing precipitation lapse rates**

In a case study, we compare CHELSA-EarthEnv's annual precipitation climatology in coastal British Columbia with that of PRISM, simulation data from the Weather Research and Forecasting (WRF) convection-permitting dynamical simulation for North America [87]; and WorldClim2.1. We calculated horizontal precipitation gradients for each grid cell by multiplying precipitation lapse rate by the terrain slope. The precipitation lapse rate is calculated from a moving window regression of precipitation against elevation in the 8 cells surrounding the focal cell.

**Accessing the improvement from the cloud layers**

We validate the inclusion of the cloud frequencies from MODIS in two steps. First, we compare the global performance of the precipitation dataset with, and without cloud refinement globally using GHCN-D. The refinement however, is done at the 0.0083334° resolution, and the mesoscale patterns of the data with, or without refinement are nearly identical. To compare the to datasets with, and without refinement at the scale where an effect of the cloud layer is actually expected, we use the island of Hawai'i as an example. Here both the station density and the quality of the stations are high, and the island has step precipitation gradients ranging from nearly 0 to >20 kg m$^{-2}$s$^{-1}$. We use 105 stations that recorded at least 25 days per month between 2003 and 2016 from GHCN-D dataset and compare the annual mean precipitation it to the one derived at the original 0.25° resolution, the data without cloud refinement, and the data with cloud refinement at 0.0083334° resolution.

**Global downscaling performance across several metrics**

Kling-Gupta Efficiency, as well as Pearson's r values were highest in Europe, Central Asia, and North America (Supplementary Fig. 1). The lowest values are found within the tropics, but also in areas with very high precipitation, such as Venezuela, Colombia, or the Congo basin, or very low precipitation, such as the Sahara, or the Arabian Peninsula. There are several possible explanations for the relatively lower performance in the tropics. We are using the GHCN-D dataset for validation, as it is one of the few available datasets for large-scale, global validation of precipitation. Gauge data such as GHCN-D is however very heterogeneous in quality [30,84–86] and, even after cleaning using the provided quality flags, errors likely remain. The lower validation performance in these regions may therefore be partially an artefact of poor station data quality.

Differences in KGE values between coarse and high resolution are higher in areas with large spatial heterogeneity such as mountains (Fig. 1). This shows that the downscaling has a positive effect on the estimation of precipitation at high spatial resolutions (Table 3). The increase in KGE values is however, not confined to areas with heterogeneous terrain, but also the lowlands in the United States or Europe. The high-resolution data shows improvements in KGE and all of its components compared to the coarse 0.25° data. Performance gains are given for the root mean squared error (*rmse*), mean absolute error (*mae*), and percent bias (*pbias*) (Fig. 1). The global performance gain is $\Delta KGE = 0.045$, but shows a strong geographical pattern (Fig. 1) especially in mountainous regions such as the Andes, or the Rocky Mountains, but also large parts of Asia. Performance losses are most prominent in Western Indonesia, with the rest of Indonesia however, showing a gain in KGE.

While globally an increase in the γ component of KGE is larger than the increase in the β or *r* component, in most of the regions with the highest gain in KGE, both increases in *r* and β prevail. A possible explanation for this is that the inclusion of topography in the downscaling has the largest effect on the bias (Fig. 1).

The more evenly distributed differences in the γ component, which reflects the variability in precipitation is most likely due to the inclusion of the MODIS cloud cover, that adds additional information on the spatio-temporal variance in precipitation to the downscaling. Although we only included monthly cloud frequency distributions into the downscaling, this shows the potential high resolution cloud cover frequencies have in improving high resolution precipitation estimates globally.

**Table 3.** Global test metrics for a comparison between the downscaled CHELSA_EarthEnv data and the original ERA5 data based on 122,236,056 observations at 58,071 stations between 2003 and 2016. *rmse* = root mean squared error, *mae* = mean absolute error, *pbias* = percent precipitation bias [88], KGE = Kling-Gupta Efficiency, *r* = Pearson product-moment correlation coefficient, β = the ratio between the mean of the simulated values and the mean of the observed ones, γ = ratio between the coefficient of variation (CV) of the simulated values to the coefficient of variation of the observed ones. Units for the *rmse* and *mae* are in kg m$^{-2}$ day$^{-1}$.

| Resolution | *rmse* | *mae* | *pbias* | KGE | Person's *r* | β | γ |
|---|---|---|---|---|---|---|---|
| 0.0083334° | 6.592 | 2.584 | -2.500 | 0.493 | 0.562 | 0.975 | 0.828 |

| 0.25° | 6.608 | 2.628 | -2.700 | 0.448 | 0.548 | 0.973 | 0.790 |
| Difference (Δ) | -0.016 | -0.044 | -0.200 | 0.045 | 0.014 | 0.002 | 0.038 |

**Regional performance**

The comparison of all five precipitation products for the continental United States, shows a relatively high performances of all datasets (Fig. 2) ranging from a correlation of $r$~0.85 (PRISM), to $r$~0.5 (CHIRPS). CHELSA_EarthEnv performs slightly worse than MSWEP in estimating daily precipitation rates, but better than CHIRPS. PRISM performs best with the highest correlations compared to GHCN-D. The performance increases for all products when monthly climatological means, instead of daily precipitation values are used, with CHELSA, CHIRPS, and MSWEP performing almost identically. PRISM still outperforms all models slightly. WorldClim shows a comparably poor performance compared to all other products during the period 2003-2016 with low correlations ($r$~0.5) and a much higher standard deviation than all other products.

All precipitation products use part of the GHCN-D stations to parametrize their algorithms. PRISM uses the daily station data directly and uses the anomalies from long term climatologies at the stations and interpolates them to achieve a gap free anomaly surface for the CAI. The achieved performance might therefore be due to the high station density in PRISM itself. CHELSA_EarthEnv uses GPCC gridded station data at 0.25° for a bias correction, therefore the algorithm cannot force the interpolation through each station location directly, which might explain the difference between PRISM and CHELSA_EarthEnv. CHIRPS uses a smaller set of stations compared to PRISM, so the difference in performance might partly be due to the less dense station network. MSWEP uses a wide variety of input sources from remote sensed data, to reanalysis data, to station data. MSWEP therefore averages out most of the errors of a single source, which leads to a relatively high performance in the resulting precipitation estimates[12]. Interestingly, WorldClim does not perform well compared the other products, despite being parameterized on a large number of stations. This might be due to errors in the parametrization of the predictors used for the long term climatologies, or uncertainties from the CAI applied on the CRU-TS data.

**Downscaling performance in relation to comparable products**

The bias compared to observations at stations is heterogenous in all different precipitation datasets. PRISM shows the lowest bias compared to GHCN-D data, while CHELSA_EarthEnv, MSWEP, and CHIRPS show similar biases (Fig. 3). WorldClim has the largest overall bias of all five comparable products.

A similar pattern emerges when the different products are compared at the 0.25° and the highest resolution. Comparing the absolute bias of the coarse resolution aggregations with the highest available resolutions shows that all different precipitation datasets have a lower absolute bias at the highest spatial resolution (Fig. 3). The amount of bias correction however varies to a large degree, with PRISM and CHELSA_EarthEnv showing the largest bias reduction, while CHIRPS and MSWEP show a slightly lower bias reduction, and WorldClim the lowest reduction. The relative smaller reduction of CHIRPS and MSWEP to CHELSA_EarthEnv and PRISM might could be due to the lower native spatial resolution (Table 2) compared to CHELSA_EarthEnv and PRISM (Table 2). However, the monthly WorldClim timeseries has the same native spatial resolution as PRISM, and still has a very low difference in absolute bias between the high and the coarse resolution, indicating poor downscaling performance. Downscaling performance also varies geographically (Fig. 4). Generally, the bias reduction is higher in mountainous regions of the western United States, and lower in the more homogenous terrain in the east. Comparing at which stations the bias is reduced (Fig. 5), shows that PRISM, CHELSA_EarthEnv, MSWEP and CHIRPS are able to reduce the absolute bias in mountainous terrain, but also in the convective regimes of the Midwest and Southwest of the United States. WorldClim only reduces the bias in the mountainous regions, but does not reduce the precipitation bias in convective regimes.

**Comparison with PRISM**

PRISM shows consistently the highest performance metrics and is therefore a suitable benchmark for a spatially explicit comparison. Overall, all precipitation datasets show similar mesoscale patterns of precipitations (Fig. 5). Marked differences are mainly apparent in the southwestern United states, where all models are comparably dryer than PRISM. Differences are also apparent in the eastern Rocky Mountains, where CHIRPS, MSWEP, and WorldClim have a considerable dry bias, but CHELSA_EarthEnv shows more similar precipitation rates as PRISM. Overall CHELSA_EarthEnv shows the lowest differences and highest correlations to PRISM (Fig. 6) ($r$=0.97, *mae*=0.20), followed by MSWEP ($r$=0.97, *mae*=0.23) and CHIRPS ($r$=0.96, *mae*=0.23). WorldClim shows the highest differences with PRISM and the lowest correlation among all products ($r$=0.95, *mae*=0.28).

**Precipitation lapse rates**

The general similarity between CHELSA-EarthEnv and PRISM (at 800m resolution) in precipitation amount and in precipitation gradients (Figure 7d,f) is notable, given that elevation-precipitation relationships in CHELSA-EarthEnv are produced by the orographic wind effect algorithm, rather than by elevational relationships to station observations as in PRISM. The WRF simulation is independent of station observations and provides further evidence that precipitation increases with elevation in this region (Figure 7h). Weaker gradients in WRF are due to the coarser (4km) grid scale, which imposes more subdued gradients of both terrain and precipitation. The strong negative gradients in WorldClim2 (Figure 7j) are due to derivation of a precipitation-elevation relationship from stations spanning the windward (low elevation stations with high precipitation) and leeward (higher elevation stations with low precipitation) sides of the mountain range. These erroneous negative gradients produce a strong underestimation of regional precipitation (Figure 7i) as they are used to extrapolate station precipitation into higher elevations (Figure 7k) that have very low station density. This case study illustrates the utility of CHELSA-EarthEnv for mountainous regions with sparse station observations: the dynamical ERA5 reanalysis provides a physically plausible regional distribution of precipitation while the orographic wind effects algorithm provides credible local elevational gradients, even in the absence of station observations.

**Improvement from the cloud layers**

The global comparison between the predicted precipitation with and without cloud cover refinement yielded in very small differences in all test metrics indicating no significant differences in global test metrics (with cloud refinement $r$=0.609, *mae*=2.404, without refinement: *r*=0.610, *mae*=2.402). The cloud cover refinement, however happens on a spatial scale, that is not necessarily captured well by a global comparison. The local comparison for the island of Hawai'i (Fig. 8) indicates that the cloud cover refinement largely acts on the local scale, where it reduces the wet bias of the interpolation without cloud cover refinement. Without the refinement the CHELSA algorithm distributes precipitation based on wind fields and boundary layer height alone. It does not distinguish areas that are usually above the clouds very well, leading to an overestimation in precipitation in these areas. Here the cloud cover refinement shows an effect, by increasing the correlation between predicted precipitation and observed precipitation, as well as decreasing the error in the estimates (Fig. 8).

**Validation results—Conclusions**

The comparison of the coarse grid resolution with the high resolution of CHELSA_EarthEnv shows that the applied downscaling is able to increase the accuracy of the precipitation predictions in several aspects and generates realistic precipitation patterns in complex terrain. The downscaling algorithm together with remotely sensed cloud cover performs equally well as other high-resolution products in predicting precipitation. The CHELSA_EarthEnv algorithm produces similar high resolution precipitation patterns as datasets that need to be informed by a high quality, dense weather station network without directly relying on stations itself. With respect to the realistic simulation of precipitation gradients in complex terrain, it also outperforms comparable high resolution global products.

## Usage Notes

Note that because of the pixel center referencing of the input GMTED2010 data the full extent of each grid as defined by the outside edges of the pixels differs from an integer value of latitude or longitude by 0.000138888888 degree (or 1/2 arc-second). Users of products based on the legacy GTOPO30 product should note that the coordinate referencing of each grid (and GMTED2010) and GTOPO30 are not the same. In GTOPO30, the integer lines of latitude and longitude fall directly on the edges of a 30-arc-second pixel. Thus, when overlaying grids with products based on GTOPO30 a slight shift of 1/2 arc-second will be observed between the edges of corresponding 30-arc-second pixels.

CHELSA_EarthEnv differs in several aspects with the already available climatological data (CHELSA V1-V2) [37] and long term downscaled CMIP5 modelled data (CHELSAcmip5ts) [39]. The main difference is the increase in temporal resolution to a daily one, compared to the other two datasets. It is similar to CHELSA V1.x in the respect that both are 'observational' datasets, while CHELSAcmip5ts is a downscaled "modelled" dataset. A value of a climate variable given a specific day or month in CHELSA_EarthEnv, or CHELSA V1.x can therefore be seen as an event which actually has been recorded, while one in the CHELSAcmip5ts dataset is only a modelled and does not represent a real observation similar to those in the forcing CMIP5 models.

## Code Availability

The code calculating the bias correction on the CHELSA V2.0 precipitation data is written in Python 2.7 and C++ (via the SAGA-GIS api). The code for the cloud cover refinement is available here: https://gitlabext.wsl.ch/karger/chelsa_earthenv. The code for the validation is available here: https://gitlabext.wsl.ch/karger/chelsa_earthenv_validation

## Acknowledgements

D.N.K. & N.E.Z. acknowledge funding from: The WSL internal grant exCHELSA, the 2019-2020 BiodivERsA joint call for research proposals, under the BiodivClim ERA-Net COFUND program, with the funding organisations Swiss National Science Foundation SNF (project: FeedBaCks, 193907), Agence nationale de la recherche (ANR-20-EBI5-0001-05), the Swedish Research Council for Sustainable Development (Formas 2020-02360), the German Research Foundation (DFG BR 1698/21-1, DFG HI 1538/16-1), and the Technology Agency of the Czech Republic (SS70010002), as well as the Swiss Data Science Projects: SPEEDMIND, and COMECO. D.N.K. acknowledges funding to the ERA-Net BiodivERsA - Belmont Forum, with the national funder Swiss National Foundation (20BD21_184131), part of the 2018 Joint call BiodivERsA-Belmont Forum call (project 'FutureWeb'), the WSL internal grant ClimEx. We thank EarthEnv project collaborators Rob Guralnick and Brian McGill for discussions preceding and intellectually benefitting the research presented here. W.J acknowledges funding from NASA grants 80NSSC17K0282, 80NSSC20K0202, and 80NSSC18K0435.

## Author contributions

D.N.K., A.W., and W.J developed the idea. A.W. produced the monthly MODIS cloud frequency layers, D.N.K and N.E.Z. developed and implemented the precipitation downscaling and bias correction algorithm, C.M. and D.N.K conducted the validation, D.N.K wrote the first version of the manuscript and all authors contributed significantly to the revision.

## Competing interests

The authors declare no conflict of interest

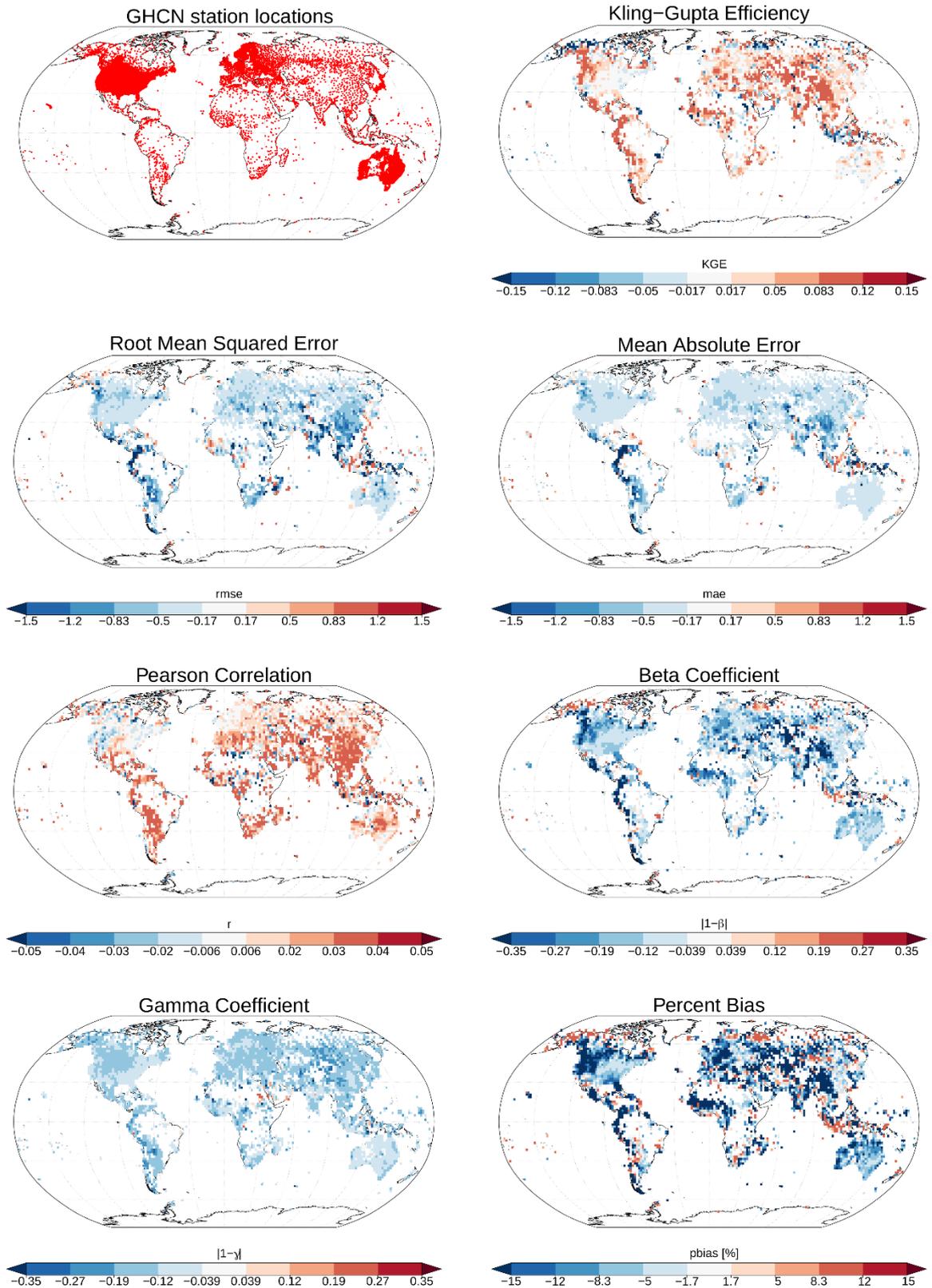

**Figure 1 | Spatial comparison of different metrics among CHELSA_EarthEnv at high resolution (0.0083334°) and coarse resolution (0.25°) based on observations from GHCN-Daily data.** From top left to bottom right: Spatial distribution of the GHCN-Daily stations without any quality issues that were used for the validation. ΔKling-Gupta-Efficiency (KGE) values between high and coarse resolution with positive values indicating an improvement in KGE in the downscaled data over the coarse grid data. Δ*rmse* and Δ*mae* values, with negative

values indicating an improvement. Differences in the correlation coefficient ($r$), with positive values indicating an improvement. $\Delta\beta$ values indicate the bias component of the KGE value, and $\Delta\gamma$ values the variance component of KGE. In both cases a negative value indicates an improvement of the downscaled values (values closer to unity). The $\Delta pbias$ gives the absolute changes in percent precipitation bias values, with negative values indicating an improvement.

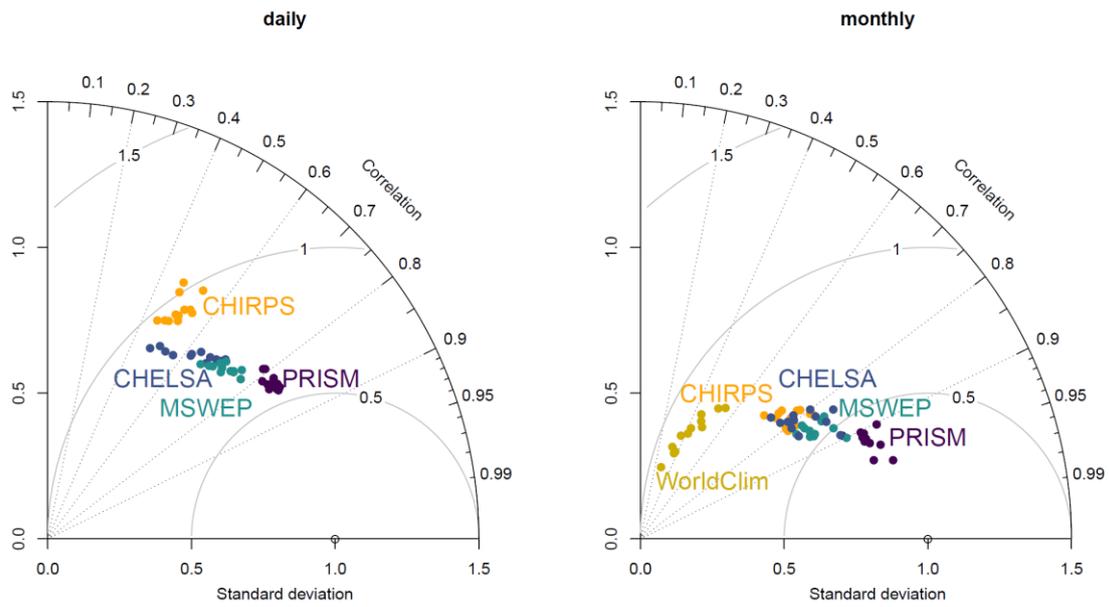

**Figure 2 | Taylor plots for comparisons between CHELSA_EarthEnv (CHELSA), PRISM, CHIRPS, MSWEP, and WorldClim with GHCN-D observations for the continental United States from 2003-2016.** The left plot is based on the comparison of daily values. Each dot represents a month. Here PRISM performs best, with MSWEP, CHELSA and CHIRPS following in that order. The plot on the right shows the performance of monthly climatological means (2003-2016). Here the aggregation of precipitation values leads to an increase in performance of all models with PRISM still showing the highest performance, with CHELSA, MSWEP, and CHIRPS performing equally well. The WorldClim monthly timeseries does not perform well with low correlation and a high standard deviation.

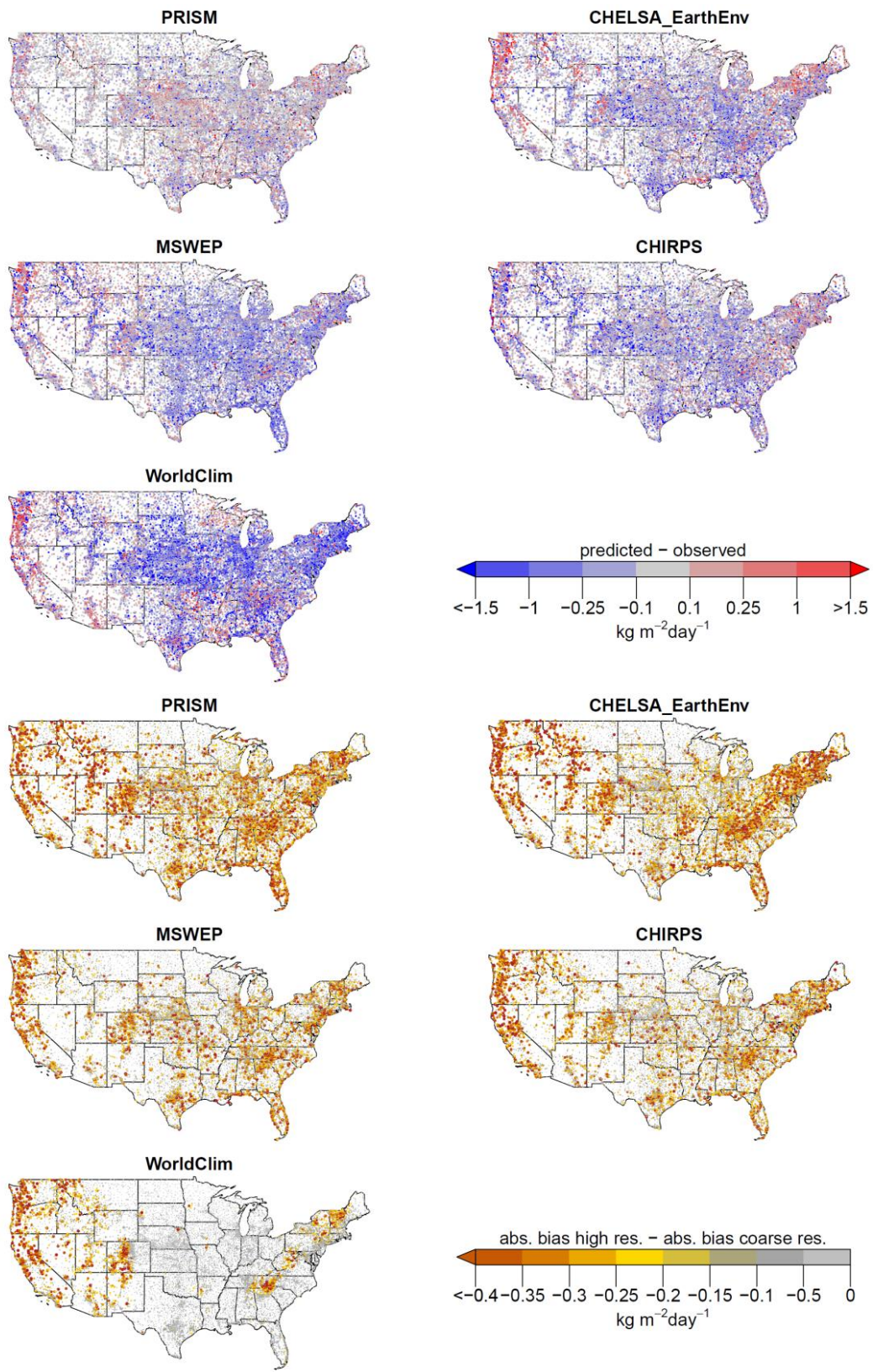

**Figure 3 | A. Bias in annual mean precipitation estimates for five different precipitation datasets compared to observations from GHCN-D.** PRISM shows the lowest bias, while all other three products have dry bias in the eastern United States. WorldClim has the highest bias of all products, with a large dry bias in the central United States. CHELSA_EarthEnv has a wet bias mainly in mountainous regions, while MSWEP and CHIRPS have a dry bias specifically

in the eastern Rocky Mountains. **B.** Changes in absolute bias between a coarse resolution (0.25°) and the native resolution of the different datasets. Negative values indicate that the downscaling decreases the absolute bias. All datasets show a lower absolute bias in mountainous terrain at a high resolution. In all datasets except for WorldClim, the higher resolution shows a lower absolute bias even in the convective regimes of the central United States. WorldClim is only able to reduce the bias in mountainous terrain, but the downscaling has no effect over most of the extent.

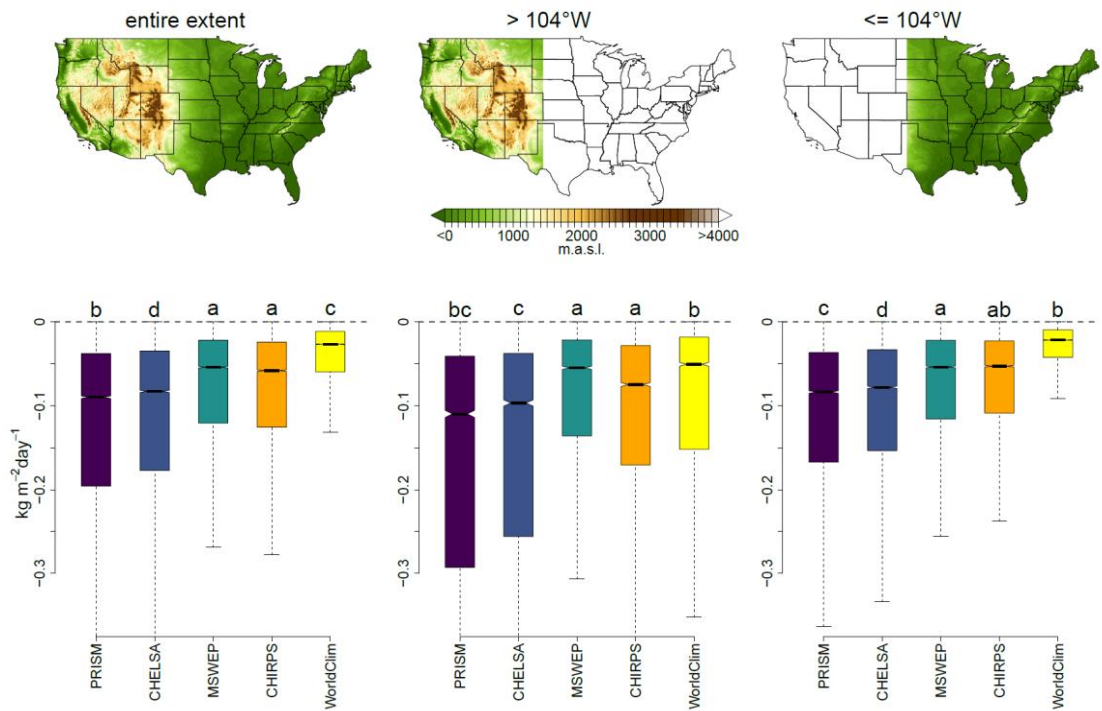

**Figure 4 | Difference in absolute bias between precipitation at 0.25° coarse resolution and precipitation at high resolution.** Data is shown for annual means and five different datasets and based on a comparison to observations from GHNC-D stations. Plots are shown separately for stations covering the entire extent of the continental United States excluding Alaska (left), stations only covering the topographic heterogeneous western part (> 104°E) (middle), and stations only covering the comparably homogeneous terrain <= 104°E in the Eastern United States (right). The more negative the value, the lower the bias of the high resolution compared to the coarse resolution. Letters are indicate a significant differences in means based on an Anova with post-hoc TukeyHSD test.

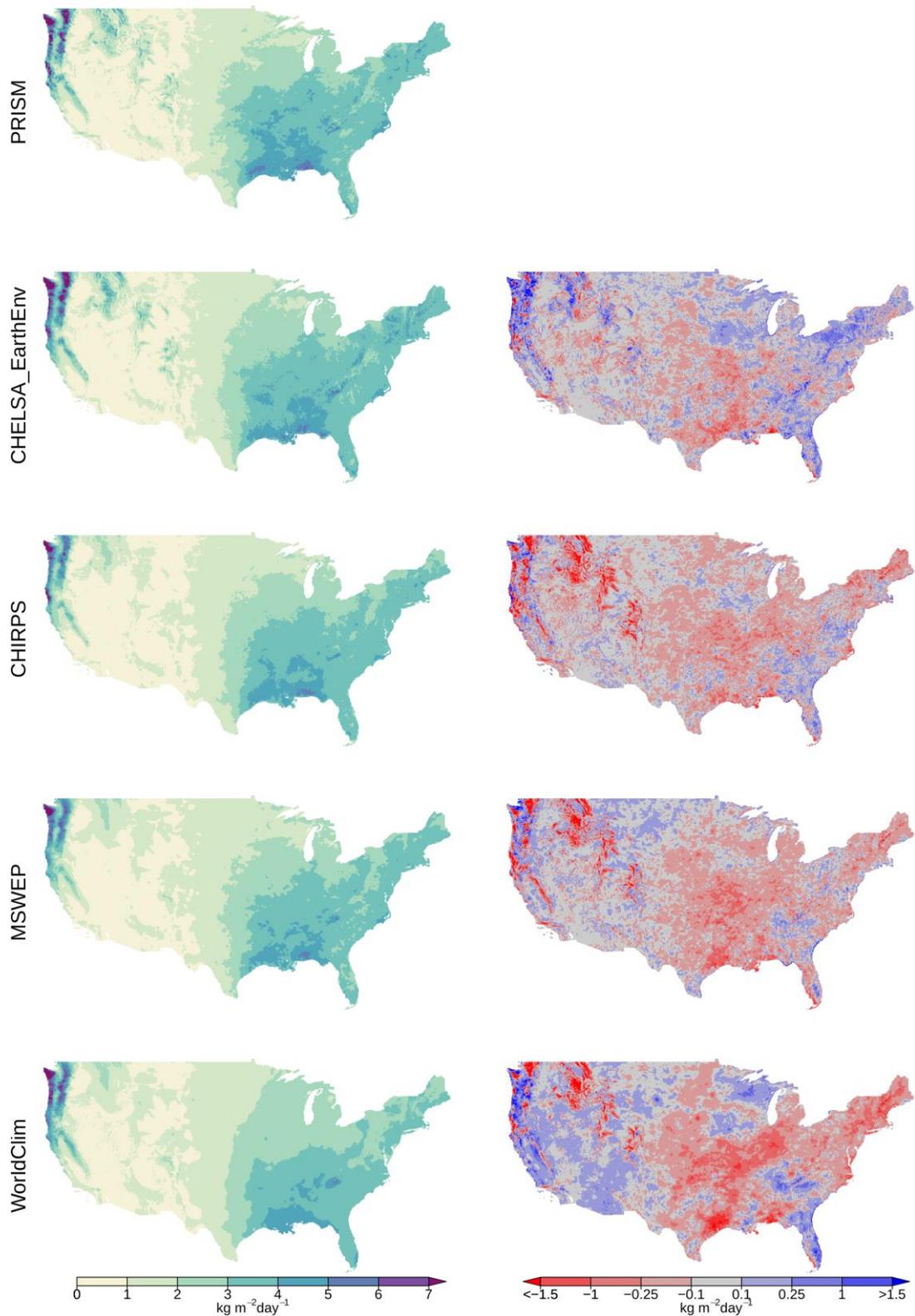

**Figure 5 | Comparison between mean annual precipitation rates (left) estimated from four different products with those of PRISM used as a reference dataset.** Annual mean of daily precipitation (left). While all three product that are compared to PRISM capture the mesoscale precipitation patterns quite well, differences exist mainly in the eastern Rocky Mountains, where CHIRPS, MSWEP and WorldClim are dryer compared to PRISM (right). CHELSA is wetter compared to PRISM mainly in mountainous terrain.

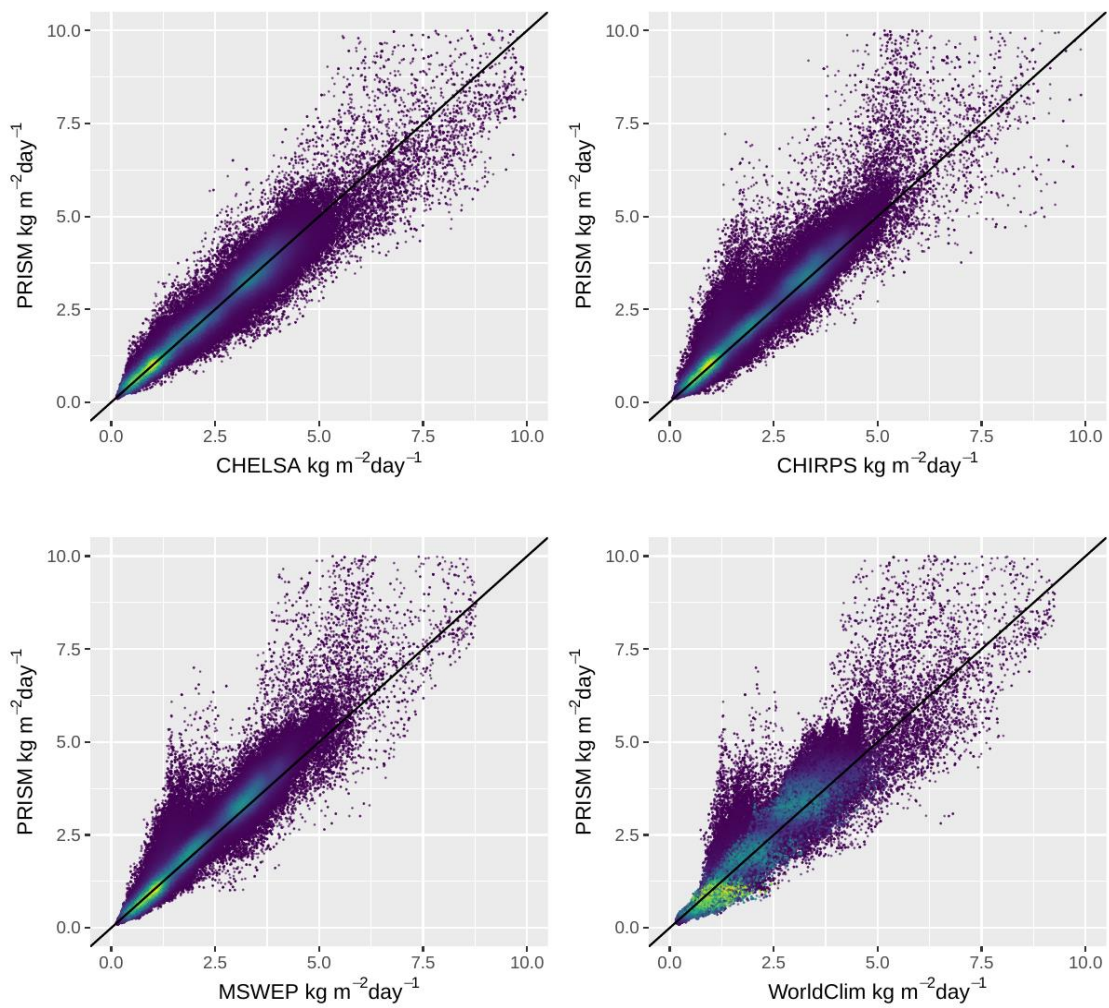

**Figure 6 | Scatterplots comparing the observed long term mean over the continental United States compared to PRISM as benchmark dataset for four precipitation datasets.** Data has been aggregated to annual means.

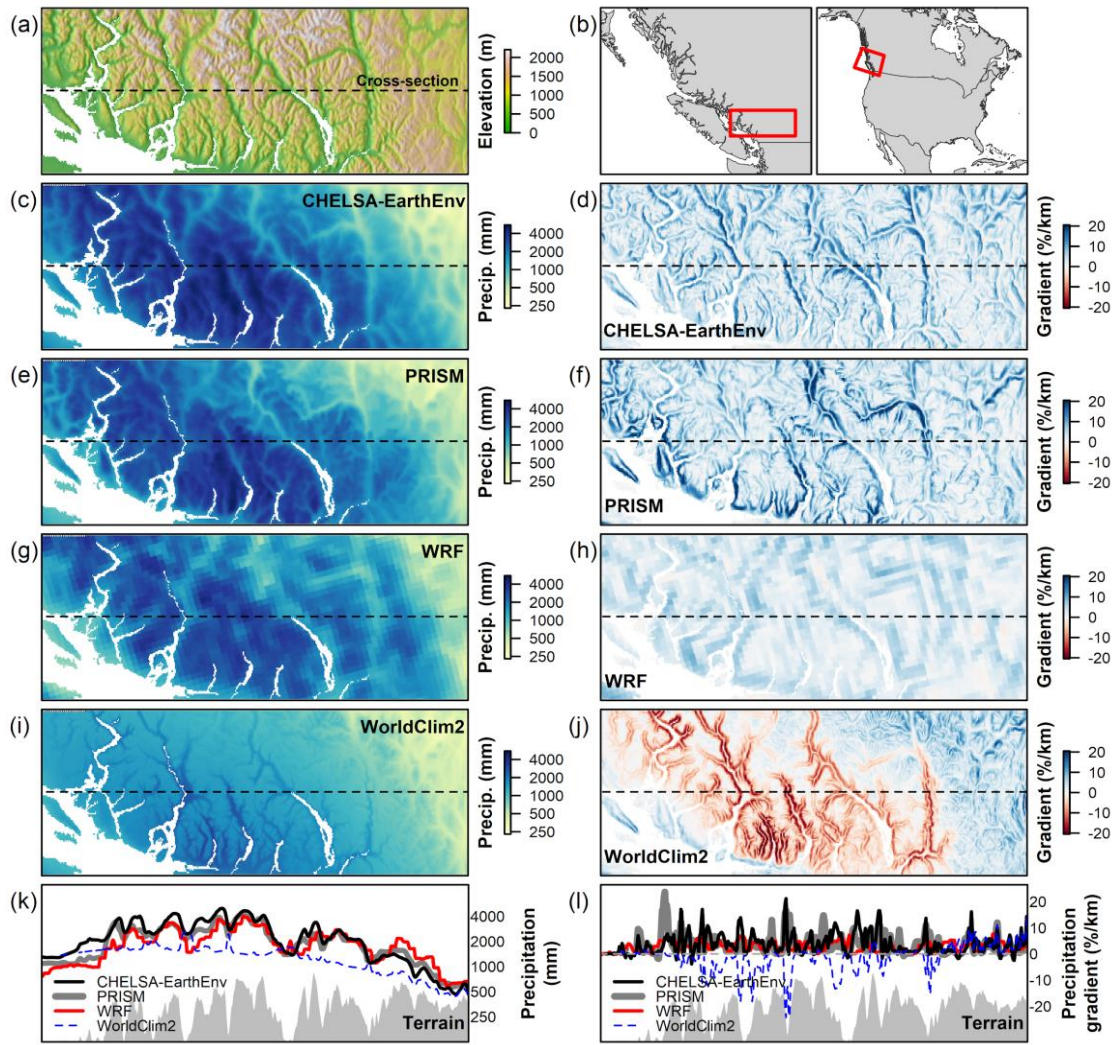

**Figure 7 | Case study intercomparison of CHELSA-EarthEnv and other gridded precipitation products for the Coast Range of British Columbia.** **(a)** Topography of the study area and location of the terrain cross-section featured in panels k-l. **(b)** location of the study area in southwestern British Columbia and North America. **(c,e,g,i)** Annual precipitation climatology for **(c)** CHELSA-EarthEnv, **(e)** PRISM, **(g)** Weather Research and Forecasting (WRF) simulation for North America, and **(i)** WorldClim2. **(d,f,h,j)** Horizontal gradients of the precipitation climatology for each product; blue (red) indicates that precipitation increases (decreases) with elevation. **(k,l)** precipitation climatologies and gradients along the terrain cross-section (gray polygon) indicated by the dashed lines in the previous panels.

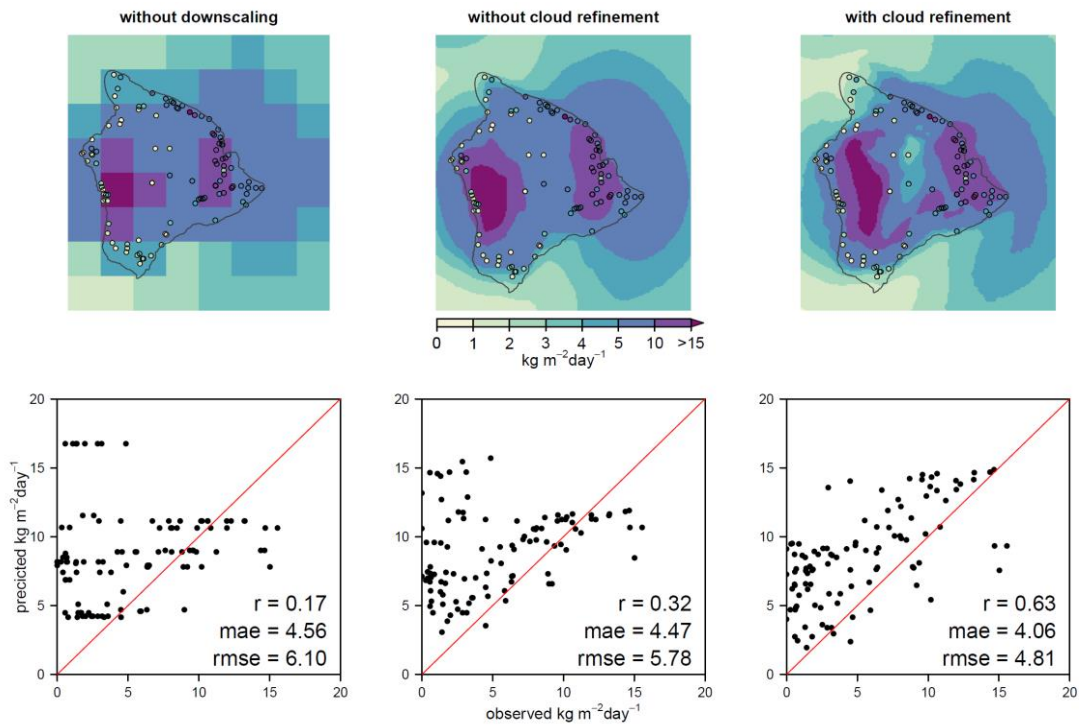

**Figure 8 | Comparison of mean annual precipitation at coarse and high-resolution, both with and without cloud refinement, for the island of Hawai'i between 2003-2016.** Points refer to observations at 105 stations from GHCN-D. While the mesoscale patterns of precipitation are relatively similar, the predictions using cloud refinement provided a stronger fit with station observations. Cloud refinement benefits prediction accuracy especially at stations with low precipitation. The Person's correlation coefficient between observed and predicted precipitation is greater at fine compared to the coarse resolution data, and for data with cloud refinement, with concomitant changes in mean absolute error (*mae*), and the root mean squared error (*rmse*). Precipitation in 2003-2016 is particularly large in the southwest of the island, an area that is usually dryer in long term (30 year) climatologies.

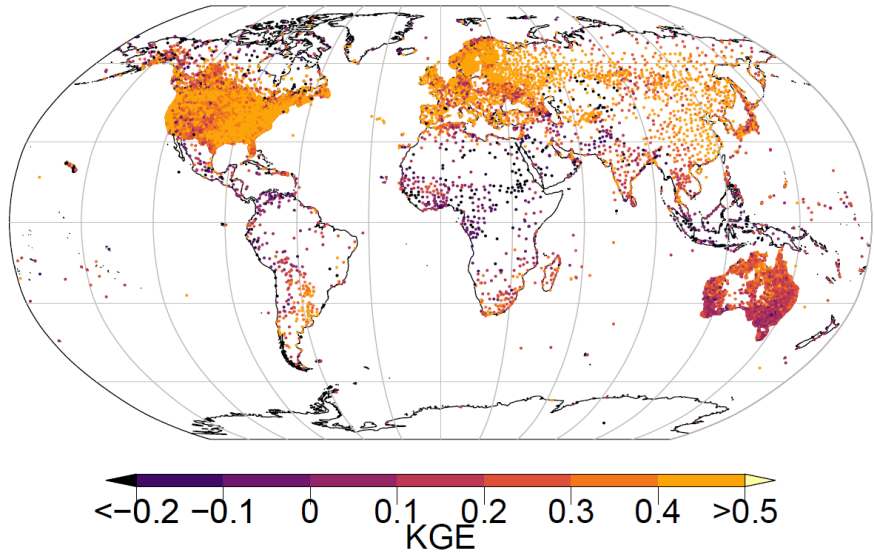

**Supplementary Figure 1 | Kling-Gupta efficiency (KGE) values derived from a comparison of CHELSA_EarthEnv with GHCN-D.** Data is based on 58,071 stations for the time period 2003 to 2016.